\documentclass{article}
\usepackage{spconf,amsmath,graphicx}
\usepackage{url,subfigure,cite,array}
\usepackage{hyperref} 
\usepackage{color}
\usepackage{booktabs}
\usepackage{tabularx}
\usepackage{multirow}

\title{Distinguishable Speaker Anonymization based on Formant and Fundamental Frequency Scaling}
\name{Jixun Yao$^1$, Qing Wang$^1$, Yi Lei$^1$, Pengcheng Guo$^1$, Lei Xie$^1$, Namin Wang$^2$, Jie Liu$^2$}
\address{$^1$Audio, Speech and Language Processing Group (ASLP@NPU)\\School of Computer Science, Northwestern Polytechnical University, Xi’an, China\\
$^2$Huawei Cloud}
\begin{document}
\ninept
\maketitle
\begin{abstract}
Speech data on the Internet are proliferating exponentially because of the emergence of social media, and the sharing of such personal data raises obvious security and privacy concerns. One solution to mitigate these concerns involves concealing speaker identities before sharing speech data, also referred to as speaker anonymization. In our previous work, we have developed an automatic speaker verification (ASV)-model-free anonymization framework to protect speaker privacy while preserving speech intelligibility. Although the framework ranked first place in VoicePrivacy 2022 challenge, the anonymization was imperfect, since the speaker distinguishability of the anonymized speech was deteriorated. To address this issue, in this paper, we directly model the formant distribution and fundamental frequency (F0) to represent speaker identity and anonymize the source speech by the uniformly scaling formant and F0. By directly scaling the formant and F0, the speaker distinguishability degradation of the anonymized speech caused by the introduction of other speakers is prevented. The experimental results demonstrate that our proposed framework can improve the speaker distinguishability and significantly outperforms our previous framework in voice distinctiveness. Furthermore, our proposed method also can trade off the privacy-utility by using different scaling factors.
\end{abstract}
\section{Introduction}
\label{sec:intro}

Speech data contains a large amount of sensitive personal information and is vulnerable to illegal exploitation by attackers. With the development of speech technology and new regulations such as the General Data Protection Regulation (GDPR) in the European Union~\cite{GDPR}, privacy preservation and protection of personal data have drawn more attention.
As a solution to speech privacy protection, speaker anonymization techniques have been developed to prevent personal speech data attacked by malicious systems.
Ideally, these techniques should preserve speech intelligibility and naturalness while suppressing speaker information, preventing the anonymized speech from being linked back to the original speaker~\cite{introprivacy}.

The task of speaker anonymization was firstly defined by the initiative called VoicePrivacy 2020 Challenge, which also introduced protocols, datasets, and metrics~\cite{introprivacy,vpc2020}. However, the challenge supposed the knowledge level of attackers was ignorant or lazy-informed, which was an improper scenario setting and reduced the difficulty of the anonymization system.
A common understanding of VoicePrivacy is still in its infancy, and the measure metrics should be involved in more aspects, e.g., voice distinctiveness and paralinguistic attributes. Therefore, the VoicePrivacy 2022 challenge~\cite{vpc2022}, as the continuation of the VoicePrivacy 2020 Challenge, focuses on concealing speaker identity to the greatest possible extent while reserving the voice distinctiveness and paralinguistic attributes intact, as well as linguistic information. In the attack model part, the knowledge level of the attacker is assumed as half-informed, which can use the anonymized training data to fine-tune the attack model.

Generally, the majority of previous anonymization systems can be broadly classified into two classes: signal processing based systems and x-vector based systems.
The signal processing based methods don't require any training data and directly modify formant, fundamental frequency, or other signal-related attributes of the speech signal to achieve anonymization~\cite{sig1,sig2,sig3}. 
These systems provide higher naturalness and are more distinguishable but less effective at protecting the speaker identity. 
However, the scope of physical shifts for speech signals is limited. One might be able to restore the original speech after a reasonable number of attempts. 
In contrast, x-vector based systems use the mean x-vector of candidate x-vectors to produce a different anonymized pseudo speaker, which achieves better privacy protection than signal processing based systems~\cite{x1,x2,x3,x4,x5,x6}. 
However, due to the introduction of other speaker information, x-vector based systems are not as well as the signal processing based systems in terms of speaker distinguishability and paralinguistic information retention. Furthermore, the x-vector based approach is difficult to flexibly trade off privacy-utility.

In the VoicePrivacy 2022 challenge, we have developed an automatic speaker verification (ASV)-model-free anonymization framework and ranked first place in four evaluation conditions~\cite{yao2022nwpu}. 
Our previous framework used a look-up table (LUT) to replace the x-vector pool and adopted a similar anonymization strategy as x-vector based approaches.
Although the developed framework is capable of the anonymization tasks, we find that the introduction of the averaged speaker embedding deteriorates the speaker distinguishability, that is, the anonymized speech from different speakers cannot be clearly distinguished. 

\begin{figure*}[ht]
        \centering
        \includegraphics[width=1.0\linewidth]{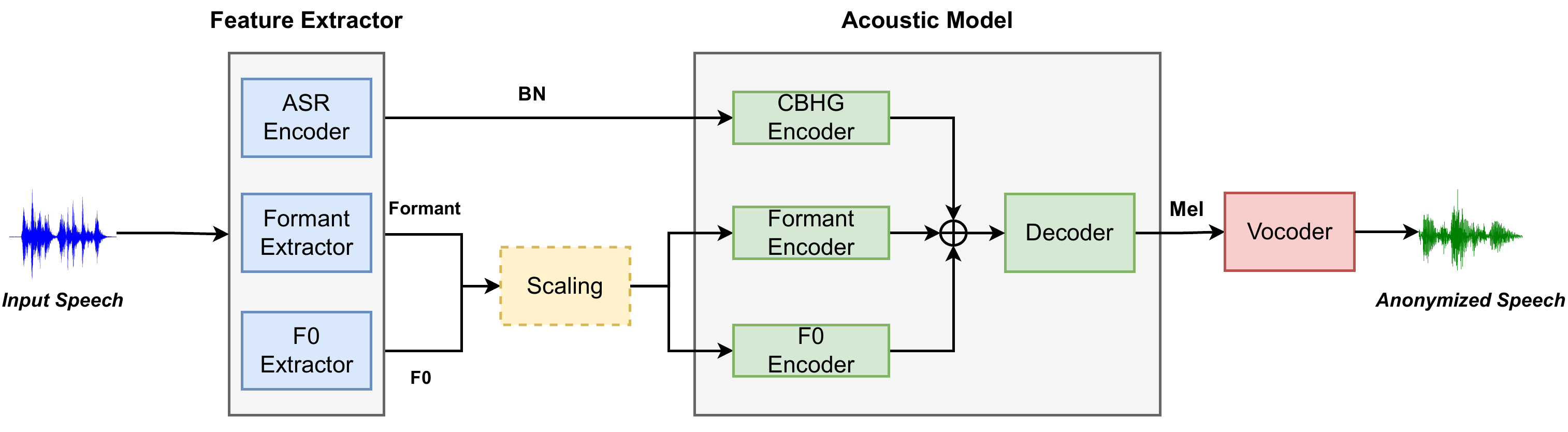}
        \vspace{-2em}
        \caption{System framework of the proposed speaker anonymization system. The dashed box is only used in the anonymized process.}
         \vspace{-10pt}
        \label{fig:model}
\end{figure*}

In this study, to tackle this speaker distinguishability problem, we propose to directly model the formant distribution and fundamental frequency (F0) from source speech that is free from additional information from other speakers. The x-vector pool or LUT is replaced by the formant and F0 to represent the speaker information. During the anonymization process, we uniformly scale the formant and F0 in each frame to suppress the speaker information, in case of being linked back to the original speaker. Moreover, the proposed framework can trade off the privacy-utility by using different scaling factors.
Experimental results show that the speech anonymized using the proposed framework significantly outperforms our previous framework in speaker distinguishability while achieving comparable results in privacy protection and speech intelligibility. 


\section{Previous Anonymization Framework}

\subsection{Overview}
In our previous work, we proposed an ASV-model-free anonymization framework~\cite{yao2022nwpu}. This framework consists of a feature extraction module, an acoustic model and a vocoder. 
The feature extraction module extracts F0, bottleneck feature (BN), and speaker ID from input speech by the YIN algorithm~\cite{de2002yin}, WeNet toolkit~\cite{wenet}
, and look-up-table (LUT), respectively. Speaker ID is utilized to generate speaker embedding by a speaker encoder and controls the speaker identity.  An encoder-decoder architecture is utilized as the acoustic model. We use the CBHG module~\cite{taco}, which does not contain an attention mechanism, as the encoder because the BN and the acoustic features are initially aligned. The decoder utilizes an autoregressive model to improve the quality and intelligibility of the reconstructed Mel-spectrogram. As for the vocoder, we modify HifiGAN with multi-band processing~\cite{hifigan, multiband}. The modified HifiGAN generator takes Mel-spectrograms as input and produces sub-band signals instead of full-band signals. In the upsampling module, 4 sub-bands are predicted simultaneously through 3 upsampling layers with 2x, 5x and 5x factors respectively, resulting in a 200x upsampling finally.

As for the anonymization strategy, the previous framework combines two types of speaker embeddings which are generated by the speaker encoder. The first one is a pseudo speaker embedding generated using a reserved pseudo speaker ID and does not correspond to any real speaker. The other one is an averaged embedding which averages the randomly selected speaker embeddings. The averaged speaker embedding and the pseudo speaker embedding are weighted concatenate to generate the anonymized speaker embedding. 
The averaged speaker embedding guarantees that each trial utterance from different speakers is anonymized by different anonymized speaker embeddings, while all trial utterances from a given speaker are anonymized by the same anonymized speaker embedding.

\subsection{Drawbacks of Our Previous Framework}
Although our previous anonymization framework ranked first place in all conditions of the VoicePrivacy 2022 Challenge, there is still an apparent drawback compared with a satisfactory anonymization system, that is, distinguishability performance is poor. 
The main reason is that each speaker has their own voice distinctiveness and the averaged speaker embedding introduced by the anonymized embedding may severely erode such distinctiveness.
What's more, the capacity of the previous framework to produce distinctive voices is also limited by the size of the look-up table and the number of speakers in the datasets.

\section{Proposed Anonymization Framework}

Since only the signal processing based method can avoid the introduction of an averaged speaker, which causes the distinguishability degradation, we propose to directly scale the formant and F0 in our distinguishability anonymization framework.
Compared with our previous framework, the proposed framework models formant and F0 by formant encoder and F0 encoder to control speaker identity instead of using LUT, without introducing any information from other speakers. 
Fig.~\ref{fig:model} shows an overview of our proposed anonymization framework.

\begin{figure}[ht]
        \centering
        \includegraphics[width=0.8\linewidth]{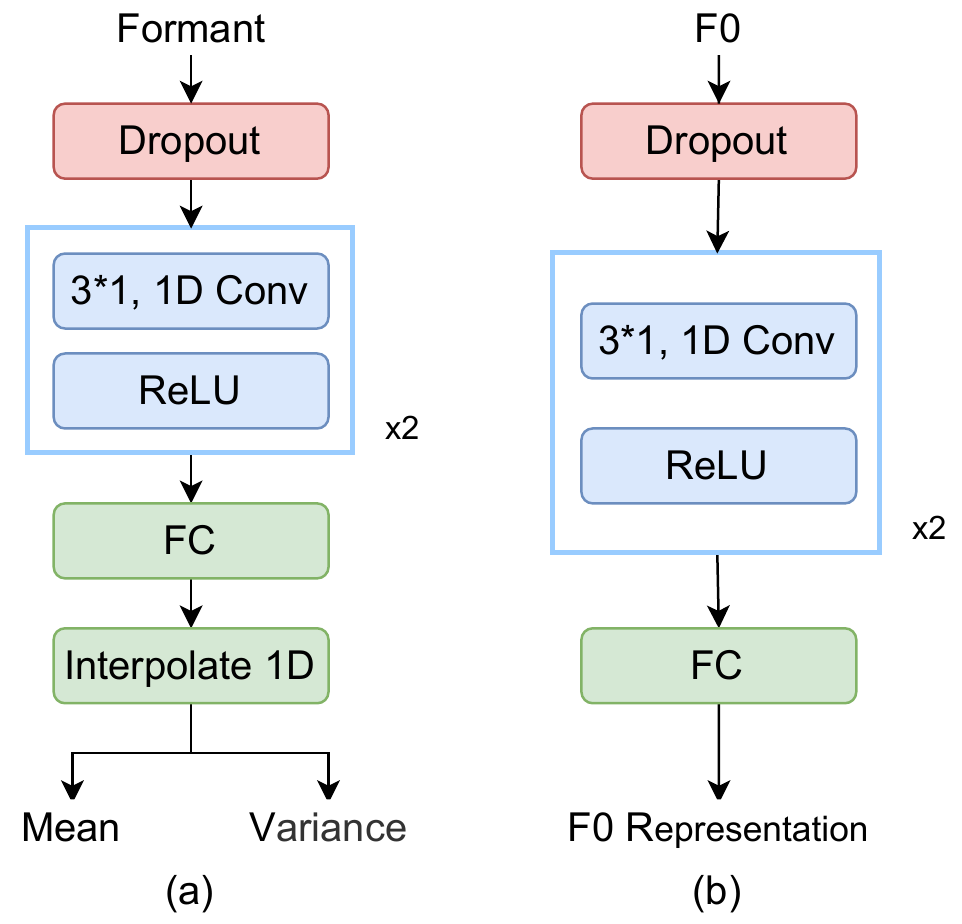}
        \vspace{-1em}
        \caption{The architectures of the formant encoder and F0 encoder: (a) formant encoder, (b) F0 encoder.}
         \vspace{-20pt}
        \label{fig:encoder}
\end{figure}

\begin{table*}[ht]
\caption{Anonymization results using the \textbf{gender-independent scaling} strategy with LibriSpeech-test and VCTK-test datasets. Various values of factor $\alpha$ scale all the formant and F0 trajectories of both genders. F, M and A denote female, male and average results, respectively.}\label{tab:exp1}
    \centering
    \renewcommand\arraystretch{1.1}
    \renewcommand{\tabcolsep}{0.1cm}
    \resizebox{1.0\linewidth}{!}{
    \begin{tabular}{c|cccccc|cccccc|cccccc|cccccc}
    \hline
    \multirow{3}{*}{\boldmath{$\alpha$}} & \multicolumn{6}{c|}{\textbf{EER (\%) $\uparrow$}}                                                & \multicolumn{6}{c|}{\textbf{WER (\%) $\downarrow$}}                                                & \multicolumn{6}{c|}{\boldmath{$\rho^{F0}$ $\uparrow$}}                                                & \multicolumn{6}{c}{\boldmath{$G_{vd}$ (close to 0)}}                                                       \\ \cline{2-25} 
                                    & \multicolumn{3}{c|}{Libri-test}        & \multicolumn{3}{c|}{VCTK-test} & \multicolumn{3}{c|}{Libri-test}        & \multicolumn{3}{c|}{VCTK-test} & \multicolumn{3}{c|}{Libri-test}      & \multicolumn{3}{c|}{VCTK-test} & \multicolumn{3}{c|}{Libri-test}             & \multicolumn{3}{c}{VCTK-test} \\ \cline{2-25} 
                                    & F     & M     & \multicolumn{1}{c|}{A} & F        & M        & A        & F     & M     & \multicolumn{1}{c|}{A} & F        & M        & A        & F    & M    & \multicolumn{1}{c|}{A} & F        & M        & A        & F      & M     & \multicolumn{1}{c|}{A}     & F        & M        & A       \\ \hline
0.5                             & 32.11 & 30.31 & \multicolumn{1}{c|}{31.21} & 38.61    & 36.53    & 37.57    & 11.68 & 13.28 & \multicolumn{1}{c|}{12.48} & 20.25    & 18.55    & 19.40    & 0.63 & 0.59 & \multicolumn{1}{c|}{0.61} & 0.59     & 0.55     & 0.57     & -9.12  & -9.76 & \multicolumn{1}{c|}{-9.44} & -10.22   & -9.8     & -10.01  \\
0.7                             & 25.99 & 23.65 & \multicolumn{1}{c|}{24.82} & 34.67    & 30.07    & 32.37    & 4.53  & 5.05  & \multicolumn{1}{c|}{4.79}  & 9.46     & 10.88    & 10.17    & 0.83 & 0.81 & \multicolumn{1}{c|}{0.82} & 0.80      & 0.76     & 0.78     & -4.65  & -5.11 & \multicolumn{1}{c|}{-4.88} & -5.12    & -5.74    & -5.43   \\
0.9                             & 20.15 & 17.05 & \multicolumn{1}{c|}{18.61} & 27.88    & 22.98    & 25.43    & 3.99  & 4.37  & \multicolumn{1}{c|}{4.18}  & 8.57     & 9.87     & 9.22     & 0.81 & 0.81 & \multicolumn{1}{c|}{0.81} & 0.81     & 0.79     & 0.80     & -3.77  & -4.25 & \multicolumn{1}{c|}{-4.01} & -4.51    & -5.21    & -4.86   \\
1.1                             & 16.52 & 19.22 & \multicolumn{1}{c|}{17.87} & 24.02    & 23.92    & 23.97    & 4.23  & 4.19  & \multicolumn{1}{c|}{4.21}  & 10.76    & 9.28     & 10.02    & 0.81 & 0.79 & \multicolumn{1}{c|}{0.80} & 0.77     & 0.79     & 0.78     & -4.10   & -4.22 & \multicolumn{1}{c|}{-4.16} & -4.98    & -4.90     & -4.94   \\
1.3                             & 22.82 & 26.94 & \multicolumn{1}{c|}{24.88} & 28.87    & 33.09    & 30.98    & 5.21  & 4.81  & \multicolumn{1}{c|}{5.01}  & 11.06    & 10.18    & 10.62    & 0.80  & 0.82 & \multicolumn{1}{c|}{0.81} & 0.78     & 0.80      & 0.79     & -5.54  & -4.66 & \multicolumn{1}{c|}{-5.10} & -5.69    & -5.13    & -5.41   \\
1.5                             & 32.88 & 33.99 & \multicolumn{1}{c|}{33.41} & 39.89    & 38.29    & 39.09    & 12.67 & 10.19 & \multicolumn{1}{c|}{11.43} & 19.22    & 18.76    & 18.99    & 0.64 & 0.70  & \multicolumn{1}{c|}{0.67} & 0.58     & 0.62     & 0.60     & -10.04 & -9.58 & \multicolumn{1}{c|}{-9.81} & -11.07   & -10.65   & -10.86  \\ \hline
\end{tabular}
}
\vspace{-15pt}
\end{table*}

\subsection{Framework Architecture}

Our proposed framework consists of three major parts: (1) a feature extraction module, (2) an acoustic model, and (3) a vocoder. 
A speech signal is anonymized by three steps. First, the feature extraction module extracts BN, formant and F0 from the original speech signal. Then, the acoustic model predicts anonymized Mel-spectrogram by BN, scaled formant and F0. Finally, the vocoder reconstructs the anonymized Mel-spectrogram to signal.

\textbf{Feature extraction}. We use the PRAAT toolkit~\footnote{https://github.com/praat/praat} to extract F1-F5 formants and F0. The purpose of the ASR encoder is to extract speaker-invariant representation. The same tool, WeNet~\cite{wenet}, is used to extract BN as in our previous work.

\textbf{Acoustic model}.
In addition to using the same CBHG encoder and autoregressive decoder as in our previous framework, the formant encoder and F0 encoder are also added to the acoustic model.
The formant encoder and F0 encoder architectures are shown in Fig.~\ref{fig:encoder} (a) and (b), they have the same architecture of a stack of two convolutional neural networks with a kernel size of 3. Differently, the output of the formant encoder is the distributions of the formant, while the high level representations are the output of the F0 encoder.
A dropout layer is also utilized at the beginning of the convolutional block to make the training process more stable. 
We adopt the L1 loss as the reconstruction loss to optimize the predicted Mel-spectrogram.

\textbf{Vocoder}. Since the output of the acoustic model is the same as the previous framework, the same vocoder and training losses are used as our previous framework.

\subsection{Anonymization Strategy}\label{sec:strategy}
Since most voice information is preserved in the F0 and formant frequency ranges~\cite{th1}, whereas linguistic information is preserved in the relative formant frequency ratios~\cite{contentvec}. 
Therefore, uniformly scaling the formant and F0 is able to suppress the original speaker information while retaining the linguistic information.

We devise two anonymization strategies for scaling the formant and F0 of the source waveform, which can effectively preserve privacy and speaker distinguishability. The first strategy is \textbf{gender-independent scaling}, which scales the formant and F0 using the same factor directly for different genders. Suppose $\alpha$ is the scaling factor, the formant and F0 from the source speech are denoted as $\mathbf{f}_{\texttt{src}}$ and $\mathbf{p}_{\texttt{src}}$, respectively. The first scaling strategy is as follows:
\begin{equation}
    \mathbf{f}_{\texttt{anon}}=\alpha \times \mathbf{f}_{\texttt{src}},
\end{equation}
\begin{equation}
    \mathbf{p}_{\texttt{anon}}=\alpha \times \mathbf{p}_{\texttt{src}},
\end{equation}
here, $\mathbf{f}_{\texttt{anon}}$ and $\mathbf{p}_{\texttt{anon}}$ are the anonymized formant and F0, whose original speaker identity is suppressed. 

The second strategy is \textbf{gender-dependent scaling}, which takes the differences between male and female in formant frequency and F0 into consideration. In general, female has higher formant frequency and F0 than male~\cite{perry2001acoustic}, so we use different scaling factors to adjust the source formant and F0 for different genders as follows:
\begin{equation}
    \mathbf{f}_{\texttt{anon}}=\left\{  
        \begin{array}{lr}  
             (1+\alpha) \times \mathbf{f}_{\texttt{src}}, \quad if\ \ male, &  \\  
             (1-\alpha) \times \mathbf{f}_{\texttt{src}}, \quad if\ \ female, &    
        \end{array}  \right.
\end{equation}
\begin{equation}
    \mathbf{p}_{\texttt{anon}}=\left\{  
        \begin{array}{lr}  
             (1+\alpha) \times \mathbf{p}_{\texttt{src}}, \quad if\ \ male, &  \\  
             (1-\alpha) \times \mathbf{p}_{\texttt{src}}, \quad if\ \ female, &    
        \end{array}  \right.
\end{equation}
The formant and F0 of the male speakers are enlarged by a factor of $(1+\alpha)$, and the female speakers are reduced by a factor of $(1-\alpha)$.
These factors enable the formant and F0 are not adjusted beyond the minimum or maximum range.
\section{Experimental Evaluations}
\subsection{Datasets}
We use the same datasets as we employed in the VoicePrivacy 2022 Challenge~\cite{vpc2022}. That is, the ASR model is trained on LibriSpeech-clean-100 and LibriSpeech-other-500 datasets~\cite{librispeech}, while the LibriTTS-clean-100 and LibriTTS-other-500 datasets~\cite{libritts} are used to train the acoustic model and vocoder. The evaluation of the systems is performed on VoicePrivacy 2022 challenge LibriSpeech-test and VCTK-test datasets~\cite{vctk}, whose details are described in~\cite{vpc2022}. 

\subsection{Evaluation Metrics}
Following the VoicePrivacy 2022 challenge configuration, the privacy-utility metrics are the equal error rate (EER) and word error rate (WER), which rely on the official ASV and ASR systems\footnote{https://github.com/Voice-Privacy-Challenge/Voice-Privacy-Challenge-2022}. The higher EER represents better privacy protection and the lower WER represents better intelligibility. 
It is essential to emphasize that anonymization should not only lead to the speaker being de-identifiable but also preserve intonation and speaker distinguishability. Thus, the intonation is measured using the Pearson correlation~\cite{pearson} between the pitch sequences of original and anonymized utterances, denoted as $\rho^{F0}$. The range of pitch correlation is from 0 to 1, and the higher is the better. 
The gain of voice distinctiveness ($G_{vd}$) metric is used to measure the preservation of speaker distinguishability, which relies on voice similarity matrices~\cite{gvd}. The voice distinctiveness remains the same when $G_{vd}=0$ and an average increase or decrease in voice distinctiveness are indicated by a gain above or below 0.

\subsection{Experimental Results}
Two anonymization strategies are proposed in Sec~\ref{sec:strategy}.
In this subsection, we investigate the effect of scaling factor for the different formant and F0 scaling strategies, respectively.
The first one is to test the validity of anonymization using the same scaling factor for different genders, and the second is to test the use of different scaling factors for different genders.

\begin{table*}[ht]
    \caption{Anonymization results compared with other baseline systems.  VPC-B1.a~\cite{vpc2022} and VPC-B1.b~\cite{vpc2022} denote two baseline systems of the VoicePrivacy 2022 Challenge, and our previous framework is denoted as Previous~\cite{yao2022nwpu}. }\label{tab:exp3}
    \centering
    \vspace{3pt}
    \resizebox{0.9\linewidth}{!}{
        \begin{tabular}{c|cc|cc|cc|cc}
        \hline
        \multirow{2}{*}{System} & \multicolumn{2}{c|}{EER (\%) $\uparrow$}        & \multicolumn{2}{c|}{WER (\%) $\downarrow$}      & \multicolumn{2}{c|}{$\rho^{F0}$ $\uparrow$}        & \multicolumn{2}{c}{$G_{vd}$ (close to 0)}           \\ \cline{2-9} 
                                & Libri-test     & VCTK-test      & Libri-test    & VCTK-test     & Libri-test    & VCTK-test     & Libri-test     & VCTK-test      \\ \hline
        VPC-B1.a~\cite{vpc2022}               & 10.48          & 13.03          & 4.75          & 11.82         & 0.78          & \textbf{0.84}          & -9.53          & -10.99         \\
        VPC-B1.b~\cite{vpc2022}               & 8.65           & 9.22           & 4.43          & 10.69         & 0.57          & 0.67          & -5.82          & -7,37          \\
        Previous~\cite{yao2022nwpu}                & 20.62          & \textbf{39.58} & \textbf{3.87} & \textbf{7.85} & 0.68          & 0.70          & -18.98         & -17.72         \\
        Proposed                & \textbf{34.91} & 35.82          & 4.34          & 9.29          & \textbf{0.79} & \textbf{0.84} & \textbf{-4.11} & \textbf{-5.73} \\ \hline
        \end{tabular}
    }
\vspace{-15pt}
\end{table*}

\subsubsection{Results of Gender-Independent Scaling}
As for the first anonymization strategy, we control the scaling factor from 0.5 to 1.5 with an interval of 0.2.
If the scaling factor is too large or too small, the intelligibility and distinguishability of the anonymized speech are severely degraded. Our selected range covers the reasonable interval between the formant and F0 enlargement and reduction. 

Table~\ref{tab:exp1} shows the effect of the scaling factor in the range of 0.5 to 1.5. With the scaling factor increasing, the EER results increase first and then decrease. The highest rates are achieved at the scaling factor of 1.5. However, the rest anonymization metrics are getting worse when the EER results are rising. 
Compared with other intervals, the change of the scaling factor from 0.7 to 0.5 and 1.3 to 1.5 brings the most significant EER raise, but the rest anonymization metrics get worse. This means the same scaling factor strategy is effective when the scaling factor is in the range of 0.7 to 1.3, while the scaling factor out of this range will lead to a degradation.

Except for the overall performance, it is also questionable whether the scaling factor will perform diversely for different genders.
When we restrict the relative scaling factor change at 0.1, e.g. scaling factor is 0.9 or 1.1, we observe that the anonymization results of male speakers achieve better performance than female speakers at the scaling factor of 1.1. On the contrary, metrics of female speakers achieve better performance than male speakers at the scaling factor of 0.9. 
We believe this is due to the difference in pitch characteristics between male and female, therefore we use different scaling factors for different genders.

\begin{figure}[ht]
        \centering
        \includegraphics[width=1.0\linewidth]{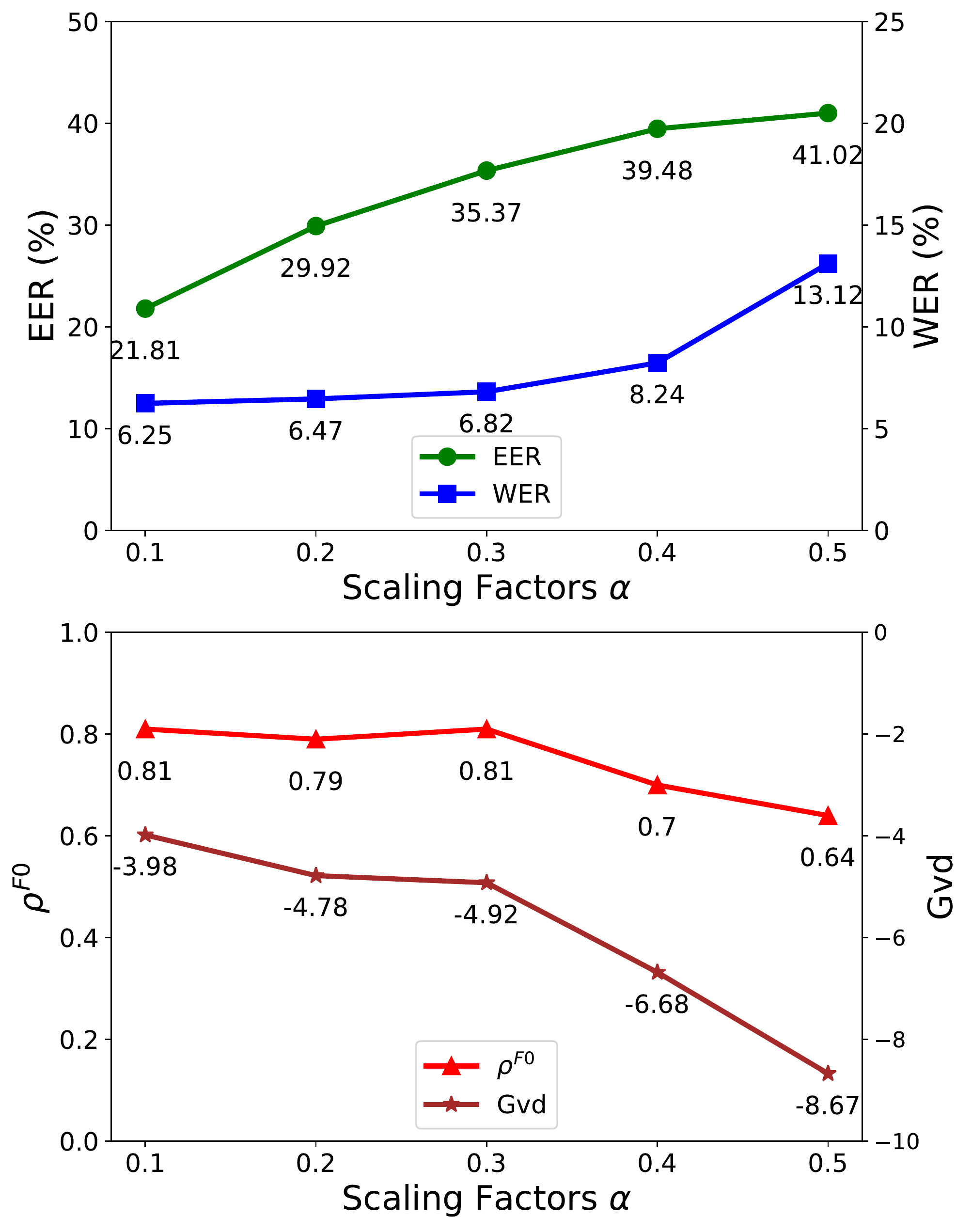}
        \vspace{-3em}
        \caption{Averaged anonymization results using the \textbf{gender-dependent scaling} strategy. The results on the LibriSpeech-test and VCTK-test datasets with equal weight. 
        }
        \label{fig:exp2}
        \vspace{-10pt}
\end{figure}

\subsubsection{Results of Gender-Dependent Scaling}
To clearly show the changes in different scaling factors,  we compute the average results of the LibriSpeech-test dataset and VCTK-test dataset for four anonymization metrics, shown in Fig~\ref{fig:exp2}. We can find the metrics show the same trend as the previous strategy but with better performance. This demonstrates that using the different scaling factors for different genders can obtain better performance than using the same scaling factor and these two strategies can trade off the privacy-utility.
On the other hand, when increasing the scaling factor from 0.1 to 0.5, the rise of the EER is getting slower and the rise of WER is getting faster, while other metrics decrease quicker and more significantly. The best trade-off result is a factor of 0.3.

\subsubsection{Comparison with Other Systems}
We configure the gender-dependent scaling strategy with a scaling factor of 0.3 to anonymize the source speech and compare it with other baseline systems. The VoicePrivacy 2022 Challenge primary baseline systems are considered as our first two baselines. Our previously submitted anonymization framework is our third baseline system which achieves the state-of-the-art privacy and utility.

As shown in Table~\ref{tab:exp3}, the EER of the proposed framework is 34.91\% and 35.82\% in LibriSpeech-test and VCTK-test datasets. The average of these two results is higher than other systems. 
We can find in $\rho^{F0}$ and $G_{vd}$ metrics that our proposed framework significantly outperforms baseline systems, especially in $G_{vd}$ metrics compared with the Previous. 
The main reason for these results is the proposed anonymization strategy does not introduce information from other speakers which deteriorates the speaker distinguishability. 
As for WER results, our framework achieves similar results with VPC-B1.a and VPC-B1.b in Librispeech-test datasets and is substantially lower than VPC-B1.a and VPC-B1.b in the WER of VCTK-test datasets. 
Although our previous framework obtains slightly lower WER results than our proposed framework, the $G_{vd}$ results show that the system is undesirable since it produces a similar voice regardless of which speaker.

\section{Conclusions}
In this study, to improve the distinguishability of speaker anonymization, instead of exploiting speaker embedding, we propose to directly model the formant distribution and F0 to represent speaker identity. The essential idea is to uniformly scale the formant and F0 to suppress the speaker identity while retaining speaker distinguishability and paralinguistic attributes, as well as linguistic information. The proposed anonymization strategy does not introduce any information from other speakers and prevents speaker distinguishability from being deteriorated. Experimental results show that the anonymization strategy can trade off privacy-utility well and significantly outperform the baseline systems in speaker distinguishability while achieving comparable privacy and utility results with the state-of-the-art anonymization system.

\bibliographystyle{IEEEbib}
\bibliography{strings,refs}

\end{document}